\documentclass{PoS}

\title{Exploratory  study of the $D_s$ spectrum in 2+1 Domain Wall QCD with heavy overlap.}

\ShortTitle{$D_s$ spectrum in 2+1 DWF with overlap charm}

\author{RBC and UKQCD Collaborations:

Chris Allton$^a$, Chris Maynard$^b$, Aurora Trivini$^a$\thanks{Speaker (E-mail: {\em pyat@swansea.ac.uk}).},   
 Robert Tweedie$^c$ \\
\llap{$^a$}Department of Physics, University of Wales Swansea, Swansea, SA2 8PP, Wales \\
\llap{$^b$}EPCC, School of Physics, University of Edinburgh, Edinburgh, EH9 3JZ, Scotland\\
\llap{$^c$}SUPA, School of Physics, University of Edinburgh, Edinburgh, EH9 3JZ, Scotland\\
}

\abstract{We present preliminary results for the $D_s$ meson spectroscopy study
on the 2+1 flavour domain wall fermion lattice configurations, generated 
with the Iwasaki gauge action at $\beta = 2.13$ by the  RBC-UKQCD collaboration. The simulations are on
 $16^3 \times 32$ lattice with $L_s = 16$.
 We consider the charm quark propagating as an overlap
fermion at fixed lattice spacing. The dispersion relation and mass splittings are evaluated. }

\FullConference{XXIVth International Symposium on Lattice Field Theory\\
                                                                                
                July 23-28, 2006\\
                                                                                
                Tucson, Arizona, USA}

\newcommand{\mc}[1]{\multicolumn{1}{c}{#1}}
\newcommand{\be}{\begin{equation} \nonumber}
\newcommand{\ee}{\end{equation}} 
\newcommand{\bea}{\begin{eqnarray} \nonumber}
\newcommand{\eea}{\end{eqnarray}}

\begin{document}

                                                                                

\section{Introduction}

The discoveries of new resonances $D_{sJ}$ by the B 
factory experiments \cite{Babar03} and CLEO \cite{Cleo03} have  provoked 
much  interest in heavy-light systems in general and in the  $D_s$ mesons
in particular. The mass splittings can be understood in terms of heavy quark and
chiral symmetry \cite{Bardeen1, B2}.

In the double limit of heavy quark and chiral symmetry, the two heavy-light
multiplets, $\{0^-,1^-\}$ and $\{0^+,1^+\}$, are degenerate.
Then chiral symmetry breaking  causes  splitting between parity partners,
such that the $1^+ - 1^-$ and $0^+ - 0^-$ are equal.
Experimentally, the splittings, shown in Table \ref{tab:expvalues}, are 
remarkably close. 
The hyperfine splitting  can also be understood in terms of heavy quark
symmetry breaking effects. 

Many previous lattice calculations \cite{PBoyle,JHein, Bali,DiPierro,CM, Ohta} tried 
to reproduce the features of these  heavy mesons, most of them considering a static or
 non-relativistic heavy charm quark, with the exception of \cite{DiPierro} which uses
 the Fermilab approach and \cite{Ohta} which describes the charm quark as
 a domain wall fermion. All these works are in the quenched approximation.

In this work  the charm quark is described by an {\em overlap}
\cite{Neuberger} formalism, while the light strange quark is a domain wall fermion, 
DWF \cite{Kaplan92}.

\begin{table}[!h]
\begin{center}
\begin{tabular}{ccc}
\mc{$0^+ - 0^-$}&\mc{$1^+ - 1^-$}&\mc{$1^- - 0^-$}\\
\hline
\hline
349.1(4) & 346.9(1.0) &  143.8(4)  \\
\end{tabular}
\end{center}
\caption{Experimental values in MeV for the different mass splittings from \cite{PDG06}.}
\label{tab:expvalues}
\end{table}



\section{Numerical details}

The gauge ensembles used for our calculations are the  2+1 flavour dynamical DWF 
 ensembles from RBC-UKQCD collaboration \cite{RBC-UKQCD}.
They were generated with the renormalized group improved Iwasaki gauge action 
at  $\beta = 2.13$. The lattice volume is $16^3\times 32$, with  the fifth dimension 
$L_s = 16$ and the domain wall height $aM_5 = 1.8$.
Three sea quark masses are considered, $am_{sea} = 0.01, 0.02, 0.03$, and 
the strange quark mass is fixed at $am_s = 0.04$ \cite{rjt}.
The correlators were measured with sources on multiple time planes, in order to 
improve our statistics.
Details of the three ensembles used are listed in  Table \ref{tab:ensembles}.

For the overlap charm \cite{Liu&Dong} quark mass, two values are chosen, 
$am_c \sim~ 0.72, ~0.9$. Correspondingly, we have two heavy-light mesons, 
indicated as H1, the lighter, and H2, the heavier, for each sea quark mass.
Recall the expression of the massive overlap operator:

\be
aD_{ov} = \rho(1+\mu)+\rho(1-\mu)\gamma_5sgn(\gamma_5(aD_W - \rho)),
\ee
 where $\mu = \frac{am_q}{2\rho}$ and $\rho$ is any mass parameter that
can be added to $D_W$ without affecting the continuum limit: here it was  chosen
equal to 1.3 looking at the heavy-heavy pseudoscalar. The overlap operator was used to
 invert on hyp-smeared DWF gauge configurations
 for mass parameter 
$\mu \sim~ 0.277, ~~0.346$, corresponding to the two charm mass values above.

\begin{table}[!h]
\begin{center}
\begin{tabular}{crccc}
\mc{$am_{sea}$}&\mc{$N_{traj}$(sep)}&\mc{No. of Origins}&\mc{$p^2_{max}$}
&\mc{$N_{cfgs}$}\\
\hline
\hline
0.01  &  500-4000(50) &  4  &  4  & 282 \\
0.02  &  1000-4025(50)&  2  &  4  & 122 \\
0.03  &  1000-4000(50)&  2  &  2  & 121 \\
\hline
\end{tabular}
\end{center}
\caption{Datasets used.}
\label{tab:ensembles}
\end{table}



\section{Analysis}

\begin{figure}[!h]
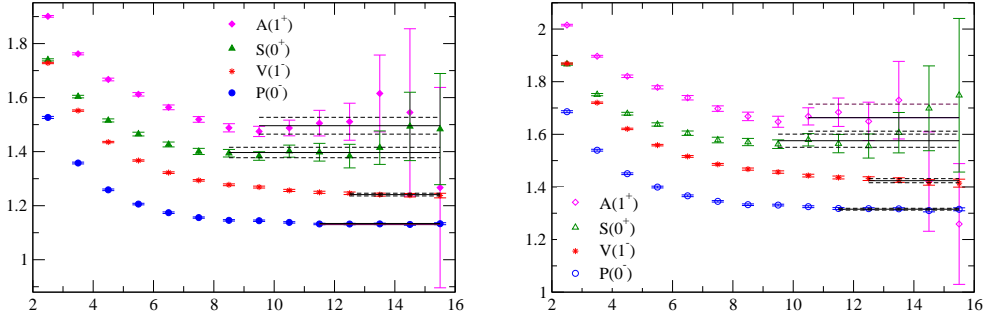

\begin{center}
\begin{minipage}[t]{0.45\textwidth}
\begin{center}
\includegraphics[width=.9\textwidth]{all4channels_h1s}
\end{center}
\end{minipage}
\begin{minipage}[t]{0.45\textwidth}
\begin{center}
\includegraphics[width=.9\textwidth]{all4channels_h2s}
\end{center}
\end{minipage}
\end{center}
\caption{Effective mass for the low-lying heavy-light mesons with charm mass 
$am_c \sim 0.72$ (left) and $am_c \sim 0.9$ (right), for the $am_{sea} = 0.01$ ensemble.}
\label{fig:effmass}
\end{figure}

In Figure \ref{fig:effmass}  we show typical effective masses for
the low-lying  $J^P$ states of the four channels we are 
interested in. The left plot is for the  heavy-light meson containing the lighter charm quark, 
the right plot for the heavier one.
For the pseudoscalar and vector channels, similarly reasonable plateau are found for higher 
momenta.
Once computed the meson masses at different lattice momenta,
\be
p_L a = \frac {2\pi\sqrt{n}}{La}, ~~~ n \in \mathbb{Z^+} 
\label{eq:p_L}
\ee
we fit them to the dispersion relation. 

The dispersion relation is  defined such that the  ${\cal O}(m^2a^2)$ 
error is reflected in the deviation of c, the effective speed of light,
from unity.
We fit the energies to a  quadratic expression as in eq. (\ref{eq:fermilab}), (\ref{eq:disp})\footnote{
 Only three momenta are available
for the $am_{sea} = 0.03$  ensemble, i.e. there are only three points in the plot of
energies versus momenta, so we found the linear fit with 
$pa = 2\sin(\pi\sqrt n/La)$ (following \cite{Liu&Dong}) be the best one 
in the 0.03 case.},
as explained below.

\begin{figure}[!h]
\begin{center}
\includegraphics[angle=0,width=0.6\textwidth]{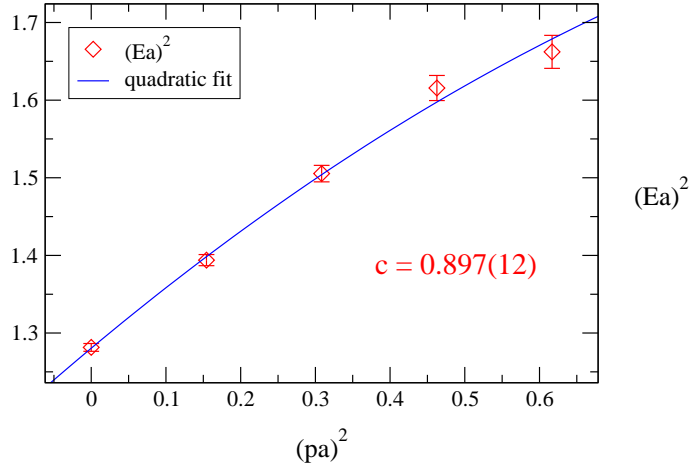}
\caption{ Dispersion relation plot  for the pseudoscalar channel 
of the lighest heavy-light meson
considered, in the $m = 0.01$ case. }
\label{fig:disp}
\end{center}
\end{figure}

The plot in Fig. \ref{fig:disp}  shows the dispersion relation for the pseudoscalar lightest
meson, i.e. H1, in the $am_{sea} = 0.01$ case with 5 momenta.
The value of the speed of light obtained from the fit to  eq. (\ref{eq:disp}),
$c = 0.897(12)$, is higher than one might expect from \cite{Liu&Dong}. 
One of the methods trying to overcome the problems with heavy quarks
(i.e. $am_Q \sim 1$) is the Fermilab or Relativistic Heavy Quark 
approach \cite{Fermilab}.
It gives us an alternative interpretation of the dispersion
relation: the basic idea is considering the expansion of the 
energy-momentum relation in powers of (lattice) momentum $pa$,
\be
(Ea)^2 = (M_1a)^2 + \frac{M_1}{M_2}{(pa)^2}+ K (pa)^4 + ...
\label{eq:fermilab}
\ee
 where $M_1$ is the rest mass, $M_1 = E(0)$, and $M_2$ is the so-called
 kinetic mass, $M_2^{-1} = (\frac{\partial^2E}{\partial{p_i^2}})_{{\bf p}=0}$.
The relativistic mass shell will have $m_Q = M_1 = M_2$, and the expression
above becomes
\be
(Ea)^2 = (M_1a)^2 + (c^2=1)(pa)^2.
\ee

In practice, at our non-relativistic mass, we can not truncate the expansion at
$p^2$, but we have to consider higher order terms, i.e. including $\delta E_{lat}$,   

\be
(Ea)^2 = (m_Qa)^2 + c^2(pa)^2  + \delta E_{lat}.
\label{eq:disp}
\ee

It has been observed that the rest mass of non-relativistic particles 
decouples from the interesting dynamics. 

The suggestion from the Fermilab approach \cite{Fermilab} is then 
considering $M_2$ instead of $M_1$ and tuning 
the couplings in the lagrangian so that $M_2$ takes the physical value.
In this preliminary analysis we consider both $M_1$ and $M_2$ and look at 
the dependence of the  mass splittings on them.


\noindent



\section{Results}

\begin{figure}
\begin{center}
\includegraphics[angle=0,width=0.5\textwidth]{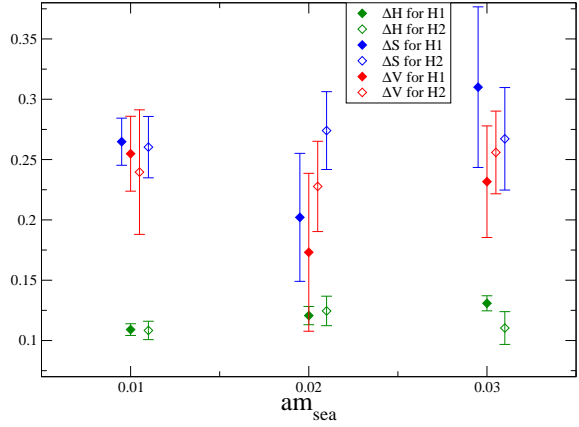}
\caption{Mass splitting values listed in Table 
 3 versus the $am_{sea}$.}
\label{fig:sea}
\end{center}
\end{figure}

First of all let's clarify the notation used for the mass splittings considered:
$\Delta H = 1^- - 0^-$ is the hyperfine splitting,  $\Delta S = 0^+ - 0^-$ and 
$\Delta V = 1^+ - 1^-$ are  the scalar and vector parity splitting respectively.
The values of these splittings obtained with all three ensembles for both
our heavy-light mesons, H1 and H2, are summarized in Table \ref{tab:splittings}.
The same splittings values are plotted versus
$am_{sea}$ in Figure \ref{fig:sea}. We can notice a very small dependence on the sea quark masses.

\begin{figure}
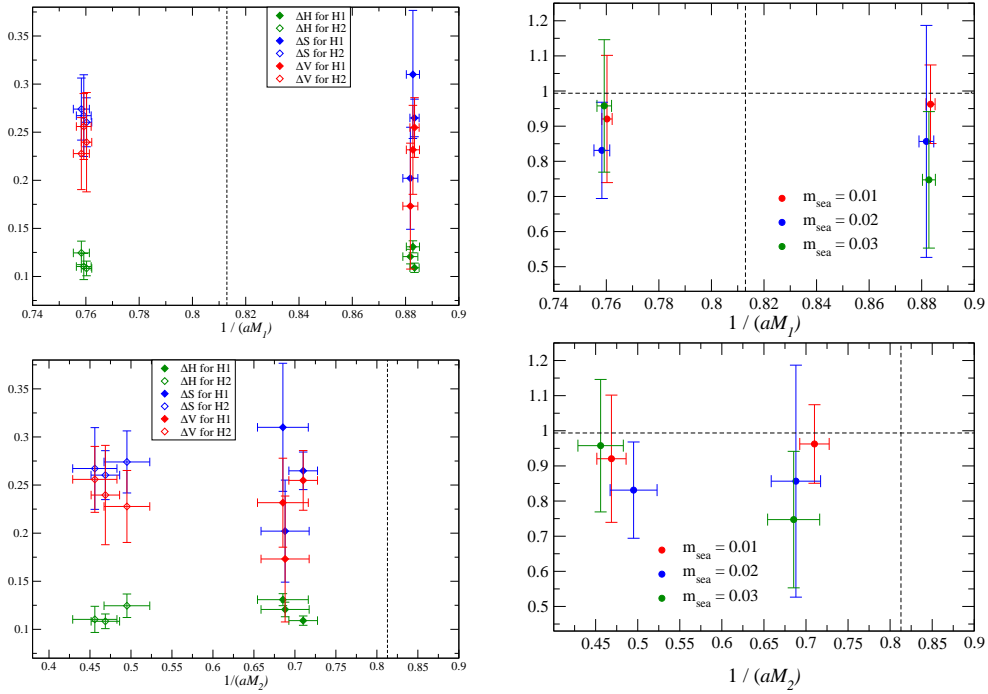

\begin{center}
\begin{minipage}[t]{0.45\textwidth}
\begin{center}
\includegraphics[width=.9\textwidth]{splittings_vs_1-M_1}
\end{center}
\end{minipage}
\begin{minipage}[t]{0.45\textwidth}
\begin{center}
\includegraphics[width=.9\textwidth]{V-S_parity_vs_1-M_PS}
\end{center}
\end{minipage}
\begin{minipage}[t]{0.45\textwidth}
\begin{center}
\includegraphics[width=.9\textwidth]{splittings_vs_1-M_2}
\end{center}
\end{minipage}
\begin{minipage}[t]{0.45\textwidth}
\begin{center}
\includegraphics[width=.9\textwidth]{V-S_parity_vs_1-M_2}
\end{center}
\end{minipage}
\end{center}
\caption{On the left the  splitting values versus $1/M_{PS}$ are plotted, with
$M_{PS}$ equal to $M_1$ in the upper left panel and to $M_2$ in the lower left one.
On the right the ratio 
$1^+- 1^- /  0^+- 0^-$ obtained is plotted 
versus $1/M_{PS}$, with $M_{PS}$ as before. The horizontal line corresponds to the
experimental value. In all plots, the vertical line represents our estimate 
of the physical $D_s$ meson, using 
$a^{-1} = 1.60(3)~Gev$ \cite{rjt}. }
\label{fig:splittings}
\end{figure}

The plots in  Figure \ref{fig:splittings} summarize our main results.
The two plots on the left show our splitting values versus $1/M_{PS}$, where
$M_{PS}$ is equal to $M_1$ in the upper left panel and to $M_2$ in the lower left one.
The two plots on the right show the ratio of vector parity over scalar parity,
$1^+- 1^- /  0^+- 0^-$ 
versus $1/M_{PS}$, as before. The horizontal line represents the experimental value.
Results from all three $m_{sea}$ values are shown. In all
plots, the vertical line represents our estimate of the physical $D_s$ meson, using 
$a^{-1} = 1.60(3)~Gev$ \cite{rjt}.
We can see that the effect of using $M_2$ instead of $M_1$ is a shift in the x axis, as
we expect from eq. (\ref{eq:fermilab}) and (\ref{eq:disp}): the difference between the
two masses is entirely a lattice artefact (eq. \ref{eq:disp}).

What we can see from these four plots is in any case no heavy quark
mass dependence of the mass splittings.


\begin{table}
\begin{center}
\begin{tabular}{cclcc}
\mc{$m_{sea}$}&\mc{Meson} &\mc{$a\Delta$H}
&\mc{$a\Delta$S}
&\mc{$a\Delta$V}\\
\hline
\hline
0.01  &   H1 & .109(5)& .265(20) &  .255(31) \\
      &   H2 & .108(8)& .260(25) &  .240(52)   \\
\hline
0.02  &   H1 & .121(8)& .202(53) & .173(65) \\
      &   H2 &  .125(12)&.274(32)& .228(37)\\
\hline
0.03  &   H1 &  .131(6) & .310(67)& .232(46)\\
      &   H2 &  .110(14)& .267(43)& .256(34)\\
\hline
\end{tabular}
\end{center}
\caption{Mass splittings value obtained for each ensemble and for both our heavy-light mesons, H1 and H2. 
 }
\label{tab:splittings}
\end{table}



\noindent




\section{Conclusions}

\begin{figure}[!h]
\begin{center}
\includegraphics[angle=0,width=0.5\textwidth]{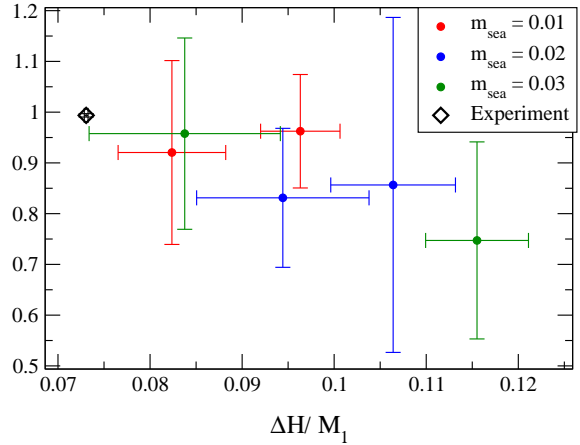}
\caption{Plot of the $1^+- 1^- /  0^+- 0^-$ ratio obtained for the three ensembles 
 versus the hyperfine splitting over $M_1$. The experimental
value is also plotted. }
\label{fig:exp_value}
\end{center}
\end{figure}

In  the first stage of our study of $D_s$ meson on 2+1 DWF QCD, with the charm quark as an
 overlap fermion, we found clear signals for all the four channels we were interested in.
A very little dependence of the splittings on the sea quark mass is observed.
For the dispersion relation analysis, both  $M_1$ and $M_2$ were considered: 
as expected,  no heavy quark mass dependance
in the splittings is observed.
The ratio of two parity splittings obtained is  close to the experimental value within statistical errors, as shown in Figure \ref{fig:exp_value}.
The $am_{sea} = 0.03$ ensemble data don't always follow the trend of the other two: investigations are in progress.
In order to reduce our large error bars, we need  more statistics.
A  quenched calculation is also in progress.


\end{document}